\documentclass[epj]{svjour}
\usepackage{graphics}
\usepackage{amsmath}
\usepackage{amssymb}

\begin{document}
\title{A note on the theory of fast money flow dynamics}
\author{Andrey Sokolov\inst{1} \and Tien Kieu\inst{1,2} \and Andrew Melatos\inst{1}
}                     
\institute{School of Physics, University of Melbourne, Parkville, VIC 3010, Australia\\ \and 
Centre for Atom Optics and Ultrafast Spectroscopy, Swinburne University of Technology, Hawthorn, VIC 3122, Australia}
\date{Received: date / Revised version: date}

\abstract{
The gauge theory of arbitrage was introduced by Ilinski in \cite{ilinski1997physics} 
and applied to fast money flows in \cite{ilinskaia1999reconcile,ilinski2001physics}.
The theory of fast money flow dynamics attempts to model the evolution of currency exchange rates and stock prices 
on short, e.g.\ intra-day, time scales.
It has been used to explain some of the heuristic trading rules, 
known as technical analysis, that are used by professional traders in the equity and foreign exchange markets.
A critique of some of the underlying assumptions of the gauge theory of arbitrage 
was presented by Sornette in \cite{sornette1998gauge}.
In this paper, we present a critique of the theory of fast money flow dynamics, which was not examined by Sornette.
We demonstrate that the choice of the input parameters used in \cite{ilinski2001physics} 
results in sinusoidal oscillations of the exchange rate, in conflict with the results presented in \cite{ilinski2001physics}.
We also find that the dynamics predicted by the theory are generally unstable in most realistic situations, 
with the exchange rate tending to zero or infinity exponentially.
} 

\maketitle

\section{Introduction}
\label{intro}
Fast money flows are analyzed in \cite{ilinskaia1999reconcile,ilinski2001physics} in terms of 
the lattice gauge theory of arbitrage developed in \cite{ilinski1997physics}.
The main idea of the theory is that the dynamics should only depend on
gauge invariant quantities rather than the exchange rates themselves.
Changing the units in which stocks of currency are denominated 
obviously changes the nominal exchange rate.
However, it is obvious that such changes of scale, i.e.~gauge transformations, should have no effect on its dynamics.
Some assumptions of the theory have been criticized in \cite{sornette1998gauge};
for example, the lack of justification for the exponential form of the weight of a given market configuration.
However, the results of the theory reported in \cite{ilinskaia1999reconcile,ilinski2001physics} seem impressive,
reproducing in particular some of the phenomenological rules of technical trading employed by professional traders.
Hence the theory appears to be a promising tool for analyzing the markets.

In this note, we present our analysis of the theory of fast money flow dynamics 
and re-examine the results presented in \cite{ilinskaia1999reconcile,ilinski2001physics}.
In Sect.~\ref{lattice}, we present the derivation of the dynamical equations of the theory.
In Sect.~\ref{analysis}, we examine the dynamics predicted by the theory for various initial conditions.
We highlight certain inconsistencies in the theory, the unstable dynamics for most realistic values of the parameters and initial conditions,
and the resulting problems in applying the theory to technical trading.
In Sect.~\ref{discussions}, we revisit the action and demonstrate that the expression used in 
\cite{ilinskaia1999reconcile,ilinski2001physics} is inconsistent with the evolution operator resulting from the lattice formulation.

\section{Lattice gauge theory and fast money flow dynamics}
\label{lattice}

In analogy with quantum electrodynamics, 
Ilinski identified the exchange rate $S$ between two currencies
with the field and the trading agents with matter.
In general, the exchange rate dynamics depends on the interest rates of the underlying currencies.
However, since we are interested in intra-day dynamics only, we consider the special case of zero interest rates.
Ilinski tacitly assumed that the interest rates of the two currencies are identical, i.e~$r_1=r_2$. 
In this paper we set $r_1=r_2=0$ and assume that transaction costs are zero.

The part of the action $s_1$ that describes the dynamics of the field on its own is formulated by
identifying arbitrage on the lattice with the curvature, which gives
\begin{equation}\label{action curvature}
s_1
=
-\frac{1}{2\sigma^2}
\int_{0}^{T}dt\,\left(\frac{dy}{dt}\right)^2.
\end{equation}
In Eq.~\eqref{action curvature}, $T$ is the investment horizon and $\sigma^2$ is the volatility 
(presumed to be constant in the interval $0\le t\le T$).
This expression is equivalent to a Gaussian random walk in $y=\ln{S}$.

The effect of the field $y$ on ``matter'', i.e.~the trading agents, is described by the Hamiltonian
\begin{equation}\label{Hamiltonian}
H(\hat\psi_1,\hat\psi_1^+,\hat\psi_2,\hat\psi_2^+)
=
H_{21}\hat\psi_1^+\hat\psi_2
+
H_{12}\hat\psi_2^+\hat\psi_1,
\end{equation}
where $\hat\psi_k^+$ and $\hat\psi_k$ are creation and annihilation operators for agents in currency $k$ ($k=1,2$),
and the coefficients $H_{21}$ and $H_{12}$ depend on  $y$.
According to Ilinski, $H_{21}=h e^{\beta y}$ and $H_{12}=h e^{-\beta y}$, where $h$ and $\beta$ are constants
(we discuss the motivation behind these formulas in Sect.~\ref{discussions}).
Following the standard treatment of a quantum harmonic oscillator (see, e.g.~\cite{slavnov1980gauge}), 
Ilinski \cite{ilinski2001physics} derived a path-integral expression
for the evolution operator in terms of the coherent states $\psi_1$ and $\psi_2$, which are the eigenstates of the 
annihilation operators $\hat\psi_1$ and $\hat\psi_2$ respectively. 
From the evolution operator one can obtain the expression for the part of the action $s_2$ that represents the field's effect on matter:
\begin{equation}
s_2
=
\int_{0}^{T}dt
\left[
\psi_1\frac{d\bar\psi_1}{dt}
+
\psi_2\frac{d\bar\psi_2}{dt}
+H(\psi_1,\bar\psi_1,\psi_2,\bar\psi_2)
\right],
\end{equation}
where the overbar denotes complex conjugation.

\begin{figure}[t!]
\centering\resizebox{1.0\columnwidth}{!}{\includegraphics{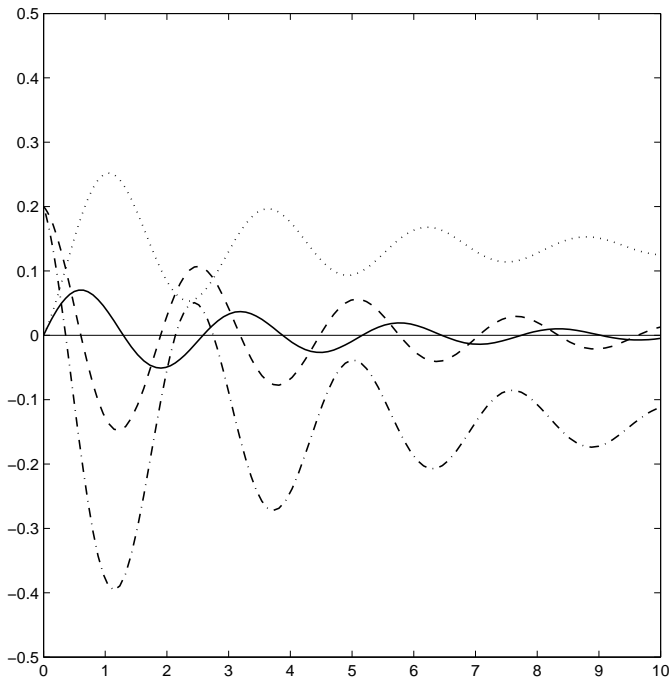}}
\caption{Re-creation of Ilinski's solution of Eqs.~(\ref{ODEeta}--\ref{ODErho})  given on page~169 of \cite{ilinski2001physics}
for $\alpha_1=1.5$, $\alpha_2=10$, $C_0=0$, and the initial conditions: $\eta(0)=0.2$, $\upsilon(0)=0$, $\rho(0)=0.5$.
The factor $\alpha_1$  in Eq.~(\ref{ODEv}) is replaced with unity to match Ilinski's Euler-Lagrange equations.
The displayed quantities are as follows:
$\rho-1/2$ (solid), 
$\upsilon+\eta$ (dashed),
$\eta$ (dot-dashed),
$\upsilon$ (dotted).}
\label{fig:ilinski}
\end{figure}
\begin{figure}[t!]
\centering\resizebox{1.0\columnwidth}{!}{\includegraphics{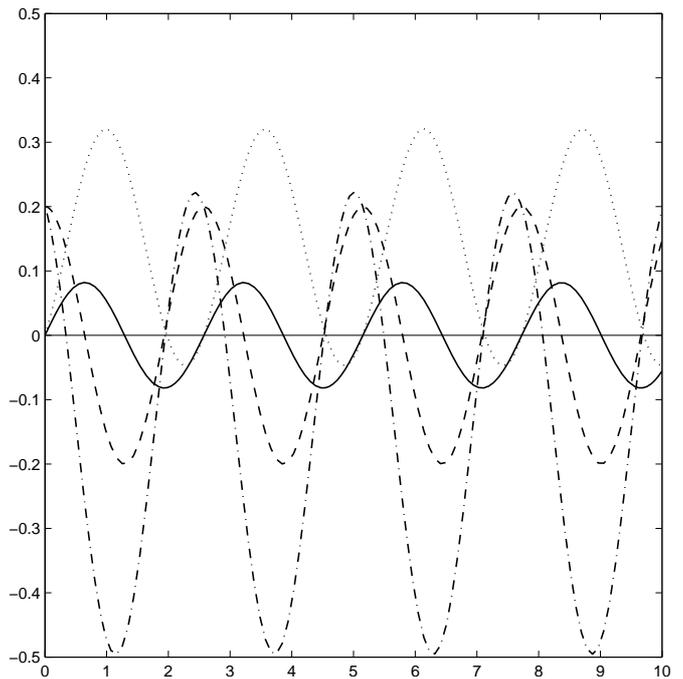}}
\caption{The solution of Eqs.~(\ref{ODEeta}--\ref{ODErho}) 
for the same parameters and initial conditions as in Fig.~\ref{fig:ilinski}.
The factor $\alpha_1$  in Eq.~(\ref{ODEv}) is restored.}
\label{fig:correct solution}
\end{figure}

Finally, departing from the electrodynamics analogy,
Ilinski introduced Farmer's term $F$ to describe the effect of matter on the field.
As a result, the action $s_1$ is replaced by
\begin{equation}
s_{1F}
=
-\frac{1}{2\sigma^2}
\int_{0}^{T}dt\,
\left[
\frac{d(y+F)}{dt}
\right]^2,
\end{equation}
where
\begin{equation}
F
=
\frac{f}{M}
(\bar\psi_1\psi_1
-\bar\psi_2\psi_2),
\end{equation}
$M$ is the total number of agents, and $f$ is a constant 
(\cite{ilinski2001physics}~uses $\alpha$ in place of $f$).

The total action is given by
\begin{equation}\label{total action}
  s=s_{1F}+s_2.
\end{equation}
Following Ilinski, we introduce new variables $\eta=\beta y$ and $\tau=ht$, 
and replace complex-valued $\psi_k$ with $\phi_k$ and $\rho_k$, defined by $\psi_k=(M\rho_k)^{1/2}e^{-i\phi_k}$ ($k=1,2$) and $\rho_1+\rho_2=1$.
Ilinski identifies $M\rho_k$ with the number of agents in currency $k$; the total number of agents is conserved.
The action can be written as
\begin{equation}
s
=
M
\int_{0}^{hT}d\tau\, L,
\end{equation}
where the Lagrangian $L$ is given by
\begin{multline}\label{lagrangian}
L
=
-(2\alpha_2)^{-1}
\left(
\eta'+\alpha_1 \rho'
\right)^2
+\rho \upsilon'
+\phi_2'
+\\
+2[\rho(1-\rho)]^{1/2}\cosh(\upsilon+\eta),
\end{multline}
with $\alpha_1=2\beta f$, $\alpha_2=M\beta^2\sigma^2/h$, $\rho=\rho_1$, $\upsilon=\phi_1-\phi_2$.
A prime denotes a derivative with respect to $\tau$.
Due to the unique structure of the Lagrangian~\eqref{lagrangian}, 
the resulting Euler-Lagrange equations can be simplified to the following first order differential equations:
\begin{align}
\label{ODEeta}
\begin{split}
\eta'
&=
 \alpha_2
(1/2-\rho)
-\\
&\quad - 2 \alpha_1 [\rho(1-\rho)]^{1/2} \sinh(\upsilon+\eta)
+ C_0,
\end{split}
\\
\label{ODEv}
\begin{split}
\upsilon'
&=
2\rho-1)[\rho(1-\rho)]^{-1/2}\cosh(\upsilon+\eta)
+\\
&\quad + 2 \alpha_1 [\rho(1-\rho)]^{1/2} \sinh(\upsilon+\eta),
\end{split}
\\
\label{ODErho}
\rho'
&=
2 [\rho(1-\rho)]^{1/2} \sinh(\upsilon+\eta).
\end{align}
However, some of the second-order nature of the Euler-Lagrange equations is retained in the constant $C_0=\eta'(0)+\alpha_1\rho'(0)+\alpha_2[\rho(0)-1/2]$,
whose value depends explicitly on the derivatives $\rho'(0)$ and $\eta'(0)$.
The equation for $\phi_2$ is trivial and we omit it.
To solve Eqs.~(\ref{ODEeta}--\ref{ODErho}), one needs to specify the initial conditions
$\eta(0)$, $\upsilon(0)$, $\rho(0)$, and $\eta'(0)$,
which uniquely determine $C_0$ (note that $\rho'(0)$ is given by Eq.~\eqref{ODErho}).
Alternatively, one can set $\eta(0)$, $\upsilon(0)$, $\rho(0)$, and $C_0$, which uniquely determine $\eta'(0)$.

\section{Analysis of the Euler-Lagrange equations}
\label{analysis}
\subsection{Missing coefficient}
\label{missing coefficient}
By introducing new variables, $\tilde\rho=\rho-1/2$ and $\tilde \eta = \upsilon+\eta$, and 
linearizing ($|\tilde\rho| \ll 1$, $|\tilde \eta| \ll1$),
we obtain $\tilde \eta=\tilde\rho'$ and
\begin{equation}
\tilde \rho''+(\alpha_2-4)\tilde\rho=C_0.
\end{equation}
For $\alpha_2>4$, the general solution is 
\begin{gather}
\tilde\rho
=
A\sin(2\pi\nu t +\theta)+C_0(\alpha_2-4)^{-1},\\
\tilde\eta
=
2\pi\nu A\cos(2\pi\nu t +\theta),
\end{gather}
with $\nu=(\alpha_2-4)^{1/2}/2\pi$ 
($A$ and $\theta$ are found from the initial conditions).
This is inconsistent with the solutions presented in~\cite{ilinskaia1999reconcile,ilinski2001physics},
which exhibit oscillations \emph{decaying slowly with time}. 
The origin of this inconsistency can be traced to a simple algebraic mistake in the derivation of the 
equations of motion given in~\cite{ilinskaia1999reconcile,ilinski2001physics}.
On page 168 of~\cite{ilinski2001physics},
the second term on the right-hand side of the equation for $\upsilon'$
is missing a factor $\alpha_1$.
The same coefficient is also missing in the equations given in~\cite{ilinskaia1999reconcile}.
This is essentially equivalent to replacing $\alpha_1$ in our Eq.~\eqref{ODEv} with unity, while keeping $\alpha_1$
in our Eq.~\eqref{ODEeta} intact.

We verify the above by numerically solving Eqs. (\ref{ODEeta}--\ref{ODErho})
in their incorrect form (with $\alpha_1$ missing from one of the equations as in \cite{ilinskaia1999reconcile,ilinski2001physics}) 
and in their correct form derived in this paper.
We are able to perfectly reproduce\footnote{In the caption of figure~7.2 in \cite{ilinski2001physics}, it is claimed 
that one of the quantities displayed is $\eta$ ($y$ in Ilinski's notation),
but actually $\eta+\upsilon$ is plotted.\label{confused tilde}}
the plots presented on page 169 of~\cite{ilinski2001physics}
by solving the incorrect equations (see Fig.~\ref{fig:ilinski}).
Note that we have $\alpha_1=1.5$ and $\alpha_2=10$ for the parameters used in \cite{ilinski2001physics}.
Ilinski claimed to set $\eta'(0)=0$ (${dy(0)}/{dt}=0$ in his notation), but this is obviously incorrect;
the solutions he presented are obtained for $C_0=0$, which gives $\eta'(0)\approx-0.3020$.
As anticipated by the linearized analysis, the correct nonlinear equations of motion
do not show any decay in the oscillation amplitude (see Fig.~\ref{fig:correct solution}).

\begin{figure}
\centering\resizebox{1.0\columnwidth}{!}{\includegraphics{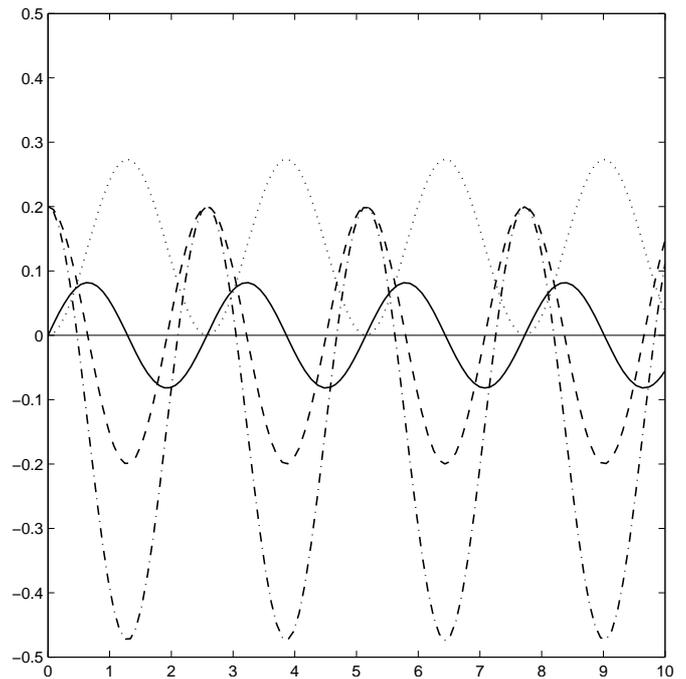}}
\caption{The solution of Eqs.~(\ref{ODEeta}--\ref{ODErho}) 
for the same parameters and initial conditions as in Fig.~\ref{fig:correct solution}, except $\alpha_1=0$.
}
\label{fig:alpha1 is zero}
\end{figure}

Furthermore, we do not observe any enhancement of oscillations for smaller values of $\alpha_1$,
as Farmer's term becomes less important.
In fact, the solutions for $\alpha_1=0$ plotted in Fig.~\ref{fig:alpha1 is zero}
are only slightly different from those for $\alpha_1=1.5$ (cf.~the plots given on page 171 of \cite{ilinski2001physics}).
After some exploration, we conclude that Farmer's term does not have any critical effect on the dynamics
of the system; it only affects the amplitude of oscillations of $\eta$ and $\upsilon$,
and their phase shift from $\rho$.

\subsection{Unstable solutions}
\label{unstable solutions}
In Sect.~\ref{missing coefficient}, we explored the dynamics of  $\tilde\eta=\eta+\upsilon$ in the case  $C_0=0$.
However, there is no a priori reason why the initial conditions should conspire to give $C_0=0$.
In this section, we briefly examine the dynamics of $\eta=\beta\ln{S}$ in the more general case $C_0\ne0$.

\begin{figure}[t!]
\centering\resizebox{1.0\columnwidth}{!}{\includegraphics{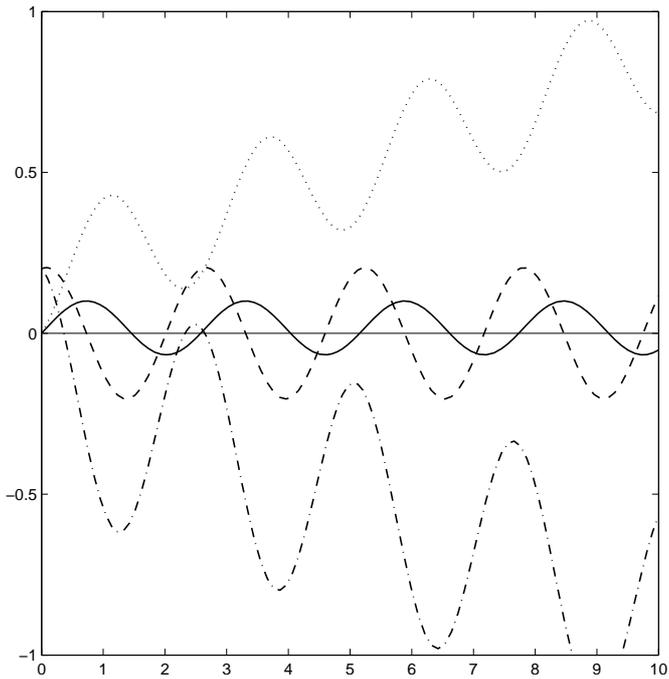}}
\caption{The solution of Eqs.~(\ref{ODEeta}--\ref{ODErho}) for the same parameters and initial conditions 
as in Fig.~\ref{fig:correct solution}, except with $C_0=0.1$ instead of $C_0=0$. 
The curves are coded as in Figs.~\ref{fig:ilinski} and~\ref{fig:correct solution}.}
\label{fig:unstable positive}
\end{figure}

\begin{figure}[t!]
\centering\resizebox{1.0\columnwidth}{!}{\includegraphics{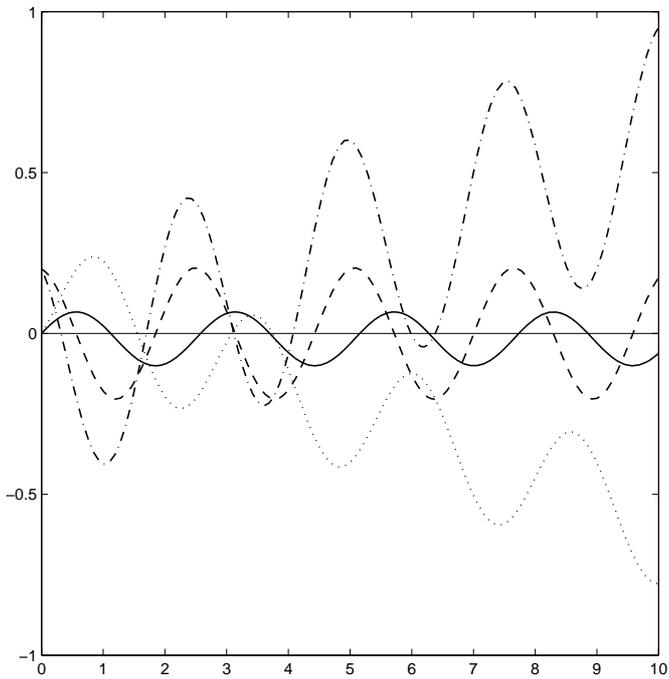}}
\caption{As for Fig.~\ref{fig:unstable positive}, with $C_0=-0.1$.}
\label{fig:unstable negative}
\end{figure}

Linearizing Eqs.~\eqref{ODEeta} and \eqref{ODEv} gives
\begin{gather}
\eta'
=-\alpha_2\tilde\rho-\alpha_1\tilde\eta +C_0,\\
\upsilon'
=4\tilde\rho+\alpha_1\tilde\eta.
\end{gather}
We find  that the solutions for $\eta$ and $\upsilon$ are also harmonic oscillations 
plus an extra term linear in time. 
The average value of $\eta$ changes linearly with time at a rate $-4C_0(\alpha_2-4)^{-1}$,
while the average of $\upsilon$ changes at the same rate but with the opposite sign.
This behaviour is illustrated in Figs.~\ref{fig:unstable positive} and \ref{fig:unstable negative}
(note that $\tilde\rho$ and $\tilde\eta$ remain small, so the linearization assumption is not broken).
Thus, for $C_0>0$, the exchange rate $S$ decays exponentially to zero,
whereas
for $C_0<0$, it grows exponentially.
In both cases the exponential time-scale is given by $\tau_c=0.25(\alpha_2-4)|C_0|^{-1}$.


\subsection{Technical trading}
\begin{figure}[h!]
\centering\resizebox{1.0\columnwidth}{!}{\includegraphics{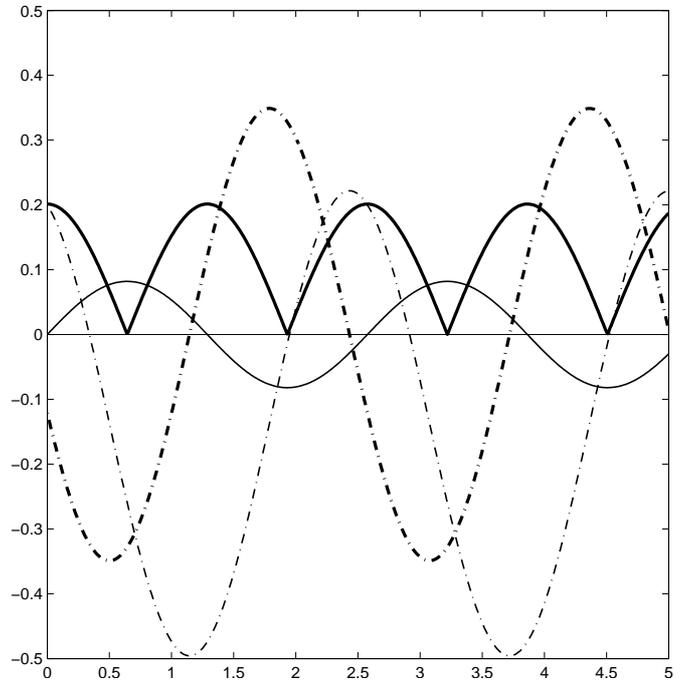}}
\caption{The trading volume $V$ (bold solid curve) and the return $R$ (bold dot-dashed curve) 
for the same parameters as in Fig.~\ref{fig:correct solution}.
For comparison, we also display $\rho-1/2$ (thin solid curve) and $\eta$ (thin dot-dashed curve).
}
\label{fig:TA}
\end{figure}

Ilinski justified certain rules employed in technical trading 
(see \cite{ilinskaia1999reconcile} and pages~170--173 of \cite{ilinski2001physics}), 
e.g., the use of positive and negative volume indices (PVI and NVI respectively), 
by appealing to the solutions of the equations of motion.
The relevant figures are presented in \cite{ilinski2001physics} on pages 170 (figure~7.3) and 172 (figure~7.7).
We identify the trading volume $V$ with $|\rho'|$ and the return $R$ with $\eta'/\beta=S'/S$.
In \cite{ilinski2001physics}, 
the derivative of $\tilde\eta=\upsilon+\eta$ is used incorrectly instead of $\eta$ to compute the return
(see also footnote~\ref{confused tilde}).
For comparison, we plot the volume and the return curves in Fig.~\ref{fig:TA},
computed using the correct equations of motion and $C_0=0$.
The quantities plotted in figure~7.7 of \cite{ilinski2001physics} are not specified, 
nor are the parameters and initial conditions, so we do not comment on that figure's validity.

Ilinski used the trading volume and the return curves 
to construct continuous\footnote{In technical trading, these quantities are discrete and defined by recursive formulas.} 
versions of PVI and NVI.
The details of the construction are left unspecified.
However, the PVI and NVI are usually computed from daily returns, not from continuous intra-day variables.
In any event, the resulting construction must
depend strongly on the time-scale that is chosen, since the indices are defined recursively.
Examining figure~7.7 in \cite{ilinski2001physics},
one observes that, for instance, the continuous PVI is constant if the trading volume $V$ is decreasing with time
and changes linearly if $V$ is increasing, with a slope of $+1$ where the return curve $R$ is positive
and $-1$ where $R$ is negative.
However, this simple trend
is inconsistent with the recursive definitions of the PVI and NVI employed in technical trading.

Moreover, the constant-amplitude solutions we employ\-ed in this section only exist for $C_0=0$.
In all other cases, the exchange rate converges to zero or diverges to infinity exponentially
on a short time scale.
The condition $C_0=0$ requires precise alignment between the initial values 
$\rho(0)$, $v(0)$, $\eta(0)$, and $\eta'(0)$.
There is no reason to expect that such a precise alignment will be observed at any time in the real market.
Therefore, the lattice gauge model predicts unrealistic behaviour (e.g., exponential divergence if $C_0<0$)
of the exchange rate under most circumstances.
Given the issues raised in this section, 
it is premature to conclude that the technical trading schemes employed by market participants
can be justified by the lattice gauge model.

\section{Revisiting the action}
\label{discussions}
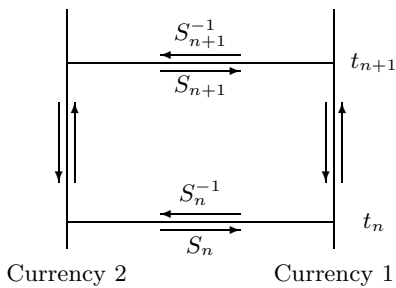
\begin{figure}[h!]
\begin{center}\footnotesize
\begin{picture}(100,100)
\put(0,10){\line(0,1){90}}
\put(100,10){\line(0,1){90}}

\put(0,20){\line(1,0){100}}
\put(0,80){\line(1,0){100}}

\put(35,17){\vector(1,0){30}}
\put(50,11){\makebox(0,0){$S_n$}}
\put(65,23){\vector(-1,0){30}}
\put(50,31){\makebox(0,0){$S_n^{-1}$}}

\put(35,77){\vector(1,0){30}}
\put(50,71){\makebox(0,0){$S_{n+1}$}}
\put(65,83){\vector(-1,0){30}}
\put(50,91){\makebox(0,0){$S_{n+1}^{-1}$}}

\put(3,35){\vector(0,1){30}}
\put(-3,65){\vector(0,-1){30}}

\put(103,35){\vector(0,1){30}}
\put(97,65){\vector(0,-1){30}}

\put(0,0){\makebox(0,0){Currency 2}}
\put(100,0){\makebox(0,0){Currency 1}}

\put(115,20){\makebox(0,0){$t_n$}}
\put(115,80){\makebox(0,0){$t_{n+1}$}}

\end{picture}
\end{center}
\caption{Lattice diagram for the intra-day foreign exchange trading in two currencies. Interest rates are ignored.}
\label{lattice diagram}
\end{figure}

We conclude by re-examining  the derivation of the action $s$ given by Eq.~\ref{total action}.
Consider two currencies, referred to as currency~1 and currency~2, 
linked by an exchange rate $S(t)$ that depends on time $t$,
such that the amount $C_2$ of currency~2 at time $t$ corresponds to the amount $C_1=S(t)C_2$ of currency~1.
We assume that the currencies can only be exchanged at the discrete times $t_n=n\Delta t$ ($n=0,\dotsc,N$) and define $S_n=S(t_n)$.
At any given time $t_n$, an agent can decide to either exchange his stock of currency for 
the counterpart currency or keep his position, in which case his stock of currency remains unchanged
(recall that we neglect interest rates completely since we are interested in the intra-day dynamics).
We display these possibilities in  Fig.~\ref{lattice diagram},
showing part of the lattice from time $t_n$ to time $t_{n+1}$.

The returns on arbitrage along the closed loops
of the elementary plaquette shown in Fig.~\ref{lattice diagram} are given by
$S_n^{-1}S_{n+1}-1$ for the clockwise loop 
and 
$S_n S_{n+1}^{-1}-1$ for the counter-clockwise loop.
The total return $S_n S_{n+1}^{-1}+S_n^{-1}S_{n+1}-2$ is identified in \cite{ilinski1997physics} with the curvature on the lattice
and, therefore, the corresponding discrete action is given by
\begin{equation}
\mathcal{A}_1
= 
\sum_{n=0}^{N}
a_n
(S_n S_{n+1}^{-1} + S_n^{-1}S_{n+1}-2).
\end{equation}
Assuming that for any $n$ we have $a_n\Delta t\to1/2\sigma^2$ in the limit $\Delta t\to0$, 
we obtain the continuous action $s_1$ given by~\eqref{action curvature}.
No justification is given in \cite{ilinski1997physics,ilinski2001physics} for why the limit of $a_n\Delta t$ must be finite.
The expression for Farmer's term was derived in \cite{ilinski2001physics}, but we omit it because its inclusion has no critical effect on the dynamics 
(see Sect.~\ref{missing coefficient}).

In order to derive the Hamiltonian given by Eq.~\eqref{Hamiltonian} and the expressions for the coefficients $H_{12}$ and $H_{21}$,
Ilinski considered the case of a single trader first and then generalized
to multiple traders by using creation and annihilation operators.
In the case of a single trader, Ilinski postulated that the probability of a given path $Q$ through
the lattice from $t_0$ to $t_N$ is exponentially weighted with respect to $s(Q)=\ln(U_1U_2\dots U_J)$,
where $\{U_j\}$ are the parallel transport coefficients on the lattice (note that $J>N$ for most paths).
Thus, for a given path $Q$, the probability is given by
\begin{equation}
P(Q)
\sim
e^{\beta s(Q)}.
\end{equation}
Depending on the path, a given  $U_j$ can be $S_n$, $S_n^{-1}$, or unity
(note that Ilinski introduces a new gauge, 
under which the exchange rates remain unchanged, 
except at $t_0$ and $t_N$ where they equal unity; see pages 131--132 of \cite{ilinski2001physics} for more details).
The state of the trader is characterized by the probabilities  $p_1$ and $p_2$ 
of being in currency 1 and currency 2 respectively.
The evolution of the state vector $(\begin{smallmatrix}p_1\\p_2\end{smallmatrix})$
can be described by the transition matrix
\begin{equation}\label{transition matrix}
P(t_n;t_{n-1})
=
\begin{pmatrix}
1& S_n^{\beta}\\
S_n^{-\beta}&1
\end{pmatrix},
\end{equation}
which Ilinski essentially identifies\footnote{
In the case of non-zero interest rates, $P(t_n;t_{n-1})$ is related to $U(t_n;t_{n-1})$
by a simple matrix transform (see page 132 of \cite{ilinski2001physics}); 
however, $P(t_n;t_{n-1})=U(t_n;t_{n-1})$ if $r_1=r_2=0$ and the transaction costs are zero.}
 with the discrete version of the continuous evolution operator $U(t,t')$ that
satisfies
\begin{equation}\label{U from H}
\frac{\partial U}{\partial t}
=
H U,
\end{equation}
where $H$ is the Hamiltonian and $U(0,0)$ is the identity matrix.
Ilinski claim that the expression for the transition matrix~\eqref{transition matrix}
and the formula~\eqref{U from H} result in
\begin{equation}\label{Hamiltonian matrix}
H=
\frac{1}{\Delta t}
\begin{pmatrix}
0&S^{\beta}\\
S^{-\beta}&0
\end{pmatrix}.
\end{equation}
Finally, identifying the parameter $h$ with $1/\Delta t$,
we obtain the expressions for $H_{12}$ and $H_{21}$, the Hamiltonian $H$ given by~\eqref{Hamiltonian}, and the action $s_2$.

In deriving the action Ilinski considered a more general case of non-zero interest rates, 
but this does not nullify the two issues pointed out below.
Firstly, we note that the Hamiltonian given by~\eqref{Hamiltonian matrix} becomes infinite in the limit $\Delta t\to0$.
It is stated in \cite{ilinski2001physics} that $\Delta t$ in the continuous-time calculations ``stands for 
the smallest time-scale of the theory, the time cut-off'' (see page~133).
However, if $\Delta t$ is retained in the finite form in the Hamiltonian and, therefore, the action $s_2$, 
it must also appear in the finite form in the expression for the action $s_1$ for consistency.
Secondly, we observe that the transition matrix $P(t_n;t_{n-1})$ is degenerate; its determinant is zero.
Therefore, it cannot possibly be identified with the evolution operator.
We conclude that the justification provided for the Hamiltonian~\eqref{Hamiltonian} in \cite{ilinski2001physics} is insufficient.

\section{Conclusions}
We have examined the theory of fast money flow dynamics developed in 
\cite{ilinskaia1999reconcile,ilinski2001physics}
and uncovered errors in 1) the derivation and the analysis of the equations of motion based
on the theory, and 2) the justification of the action based on the lattice gauge formalism.

The equations of motion presented in \cite{ilinskaia1999reconcile,ilinski2001physics}
are missing the coefficient $\alpha_1$ in one term,
crucially modifying the dynamics of the system.
We also find that most of the solutions of the equations of motion, in their correct form derived in this paper, 
are unstable with respect to the initial conditions, resulting in unrealistic behaviour of the exchange rate.
We show that the justification of the technical trading given in \cite{ilinski2001physics}
is based on an erroneous interpretation of the variables related to the exchange rate
and on the stability predicted by the incorrect equations of motion.

The theory of fast money flows relies on a particular form of the Hamiltonian that 
describes the effect of the exchange rate on the actions of the agents.
We demonstrate that this form is not consistent with the lattice gauge formulation
and diverges in the continuum limit.

\section*{Acknowledgement}

AS thanks the Portland House Foundation for their generous financial support.

\bibliographystyle{epj.bst}
\bibliography{asokolov.bib}

\end{document}